\documentclass{aa}
\usepackage{graphicx}
\usepackage{natbib}
\usepackage{txfonts}
\begin{document}
   \title{Supernovae observations and cosmic topology}

   \author{M. J. Rebou\c{c}as\inst{1}, J. S. Alcaniz\inst{2},
   B. Mota\inst{1}, \and M. Makler\inst{1}}

   \institute{Centro Brasileiro de Pesquisas F\'{\i}sicas,
              Rua Dr.\ Xavier Sigaud 150,
              CEP 22290-180, Rio de Janeiro -- RJ, Brazil \\
              \email{reboucas@cbpf.br, brunom@cbpf.br, martin@cbpf.br}
         \and
              Observat\'orio Nacional,
              Rua Gal.\ Jos\'e Cristino 77,
              CEP 20921-400, Rio de Janeiro -- RJ, Brazil \\
              \email{alcaniz@on.br}
             }

\authorrunning{Rebou\c cas, Alcaniz, Mota \& Makler}
\titlerunning{Supernovae observations and cosmic topology}



  \abstract
{Two fundamental questions regarding our description of the Universe
concern the geometry and topology of its $3$-dimensional space.
While geometry is a local characteristic that gives the intrinsic
curvature, topology is a global feature that characterizes the shape
and size of the $3$-space. The geometry constrains, but does not
dictate, the spatial topology.}
{We show that besides determining the spatial geometry, the
knowledge of the spatial topology allows us to place tight
constraints on the density parameters associated with dark matter
($\Omega_m$) and dark energy ($\Omega_{\Lambda}$).}
{By using the Poincar\'e dodecahedral space as the observable
spatial topology, we reanalyze the current type Ia supernovae (SNe
Ia) constraints on the density parametric space $\Omega_{m} -
\Omega_{\Lambda}$.}
{From this SNe Ia plus cosmic topology analysis, we find best-fit
values for the density parameters that are in agreement with a
number of independent cosmological observations.}
   {}

   \maketitle
%

\section{Introduction}

Cosmologists assume that the Universe can be described as a
manifold. Mathematicians characterize manifolds in terms of their
geometry and topology. Thus, two fundamental questions regarding our
understanding of the Universe concern its geometry and topology. An
important difference between these two attributes is that while
geometry is a local characteristic that gives the intrinsic
curvature of a manifold, topology is a global feature that
characterizes its shape and size.

Within the framework of standard cosmology, the Universe is
described by a space-time manifold $\mathcal{M}_4 = \mathbb{R}
\times M$ with a locally homogeneous and isotropic Robertson-Walker
(RW) metric
\begin{equation}
\label{RWmetric} ds^2 = -dt^2 + a^2 (t) \left [ d \chi^2 + f^2(\chi)
(d\theta^2 + \sin^2 \theta  d\phi^2) \right ] \;,
\end{equation}
where $f(\chi)=(\chi\,$, $\sin\chi$, or $\sinh\chi)$, depending on
the sign of the constant spatial curvature ($k=0,1,-1$). The spatial
section $M$ is usually taken to be one of the simply connected
spaces, namely, Euclidean $\mathbb{R}^3$, spherical $\mathbb{S}^3$,
or hyperbolic $\mathbb{H}^3$. However, this is an assumption that
has led to a common misconception that the curvature $k$ of $M$ is
all one needs to decide whether the spatial section is finite or
not.

In a spatially homogeneous and isotropic Universe, for instance, the
geometry, and therefore the corresponding curvature of the spatial
sections $M$, is determined by the total matter-energy density
$\Omega_{\mathrm{tot}}$. This means that the geometry or the
curvature of $M$ is observable, i.e. for $\Omega_{\mathrm{tot}} < 1$
the spatial section is negatively curved ($k=-1$), for
$\Omega_{\mathrm{tot}} = 1$ it is flat ($k=0$), while for
$\Omega_{\mathrm{tot}} > 1$ $M$ is positively curved ($k=1$). In
consequence, a key point in the search for the (spatial) geometry of
the Universe is to use observations to constrain the density
$\Omega_{\mathrm{tot}}$. In the context of the standard $\Lambda$CDM
model (which we adopt in this work), this amounts to determining
regions in the $\Omega_{\Lambda} - \Omega_{m}$ parametric plane that
consistently account for the observations, and from which one
expects to deduce the geometry of the Universe. As a matter of fact,
the resulting regions in this parametric plane also give information
on the dynamics of the Universe as, for example, whether an
accelerated expansion is indicated by the observations, and on the
possible behaviors regarding the expansion history of the Universe
(eternal expansion, recollapse, bounce, etc.).

However, geometry constrains, but does not dictate, the topology of
the $3$-manifold $M$. Indeed, for the Euclidean geometry ($k=0$)
besides $\mathbb{R}^{3}$, there are 17 classes of topologically
distinct spaces $M$ that can be endowed with this geometry, while
for both the spherical ($k=1$) and hyperbolic ($k=-1$) geometries
there is an infinite number of topologically inequivalent manifolds
with non-trivial topology that admit these geometries.

Over the past few years, distinct approaches to probe a non-trivial
topology of the Universe,%
\footnote{In this article, in line with the usage in the literature,
by topology of the Universe we mean the topology of the space-like
section $M$.}
using either discrete cosmic sources or cosmic microwave background
radiation (CMBR), have been suggested \citep[see, e.g., the review
articles of][]{Lach1995,Satrk1998,Levin2002,RG2004,BSCG}. An
immediate observational consequence of a detectable non-trivial
topology of the $3$-space $M$ is that the sky will show multiple
(topological) images of either cosmic objects or specific spots of
CMBR \citep{GRT2001a,GRT2001b,WLU2003,Weeks2003}. The so-called
``circles-in-the-sky" method \citep{CSS1998}, for instance, relies
on multiple images of correlated circles in the CMBR maps. In a
space with a detectable non-trivial topology, the sphere of last
scattering intersects some of its topological images along the
circles-in-the-sky, i.e., pairs of matching circles of equal radii,
centered at different points on the last scattering sphere (LSS),
with the same distribution (up to a phase) of temperature
fluctuations, $\delta T$, along the correlated circles. Since the
mapping from the last scattering surface to the night-sky sphere
preserves circles \citep{CGMR05}, the correlated circles will be
written in the CMBR anisotropy maps regardless of the background
geometry and for any non-trivial detectable topology. As a
consequence, to observationally probe a non-trivial topology, one
should scrutinize the full-sky CMBR maps to extract the correlated
circles, whose angular radii, matching phase, and relative position
of their centers can be used to determine the topology of the
Universe. Thus, a non-trivial cosmic topology is an observable and
can be probed for all locally homogeneous and isotropic geometries,
without any assumption concerning the cosmological density
parameters.

In this regard, the question as to whether one can use this
observable to either determine the geometry or set constraints on
the density parameters naturally arises. Regarding the geometry it
is well-known that the topology of $M$ determines the sign of its
curvature \citep[see, e.g.,][]{BernshteinShvartsman1980}. Thus, the
topology of the spatial section of the Universe dictates its
geometry. At first sight, this seems to indicate that the bounds on
the density parameters $\Omega_{m}$  and $\Omega_{\Lambda}$ arising
from the detection of cosmic topology should be very weak, in the
sense that they would only determine whether the density parameters
of the Universe take values in the regions below, above, or on the
flat line $\Omega_{\mathrm{tot}} = \Omega_{\Lambda} + \Omega_{m}=1$.

In this article, however, we  show that, contrary to this
indication, the detection of the cosmic topology through the
``circles-in-the-sky" method gives rise to very tight constraints on
the density parameters. To this end, we use the Poincar\'e
dodecahedral space as the observable topology of the spatial
sections of the Universe to reanalyze the current SNe Ia constraints
on the parametric space $\Omega_{m} - \Omega_{\Lambda}$, as provided
by the so-called \emph{gold} sample of 157 SNe Ia given by
\citet{rnew}. As a result, we show that the knowledge of cosmic
topology provides very strong and complementary constraints on the
region of the density parametric plane allowed by SNe Ia
observations, drastically reducing the inherent degeneracies of
current SNe Ia measurements.

\section{SNe Ia observations and cosmic topology}

The value of the total density $\Omega_{\mathrm{tot}}=1.02 \pm\,
0.02$ reported by the WMAP team \citep{WMAP-Spergel}, which favors a
positively curved Universe, and the low power measured by WMAP for
the CMBR quadrupole ($\ell=2$) and octopole ($\ell=3$) moments, have
motivated the suggestion by \citet{Poincare} of the Poincar\'e
dodecahedral space topology as a possible explanation for the
anomalous power of these low multipoles. They found that the power
spectrum of the Poincar\'e dodecahedral space's fits the
WMAP-observed small power of the low multipoles, for
$\Omega_{\mathrm{tot}}\simeq 1.013$, which clearly falls within the
interval suggested by WMAP. Since then, the dodecahedral space has
been examined in various studies
\citep{Cornish,Roukema,Aurich1,Gundermann,Aurich2}, in which further
features of the model have been carefully considered. As a result,
it turns out that a Universe with the Poincar\'e dodecahedral space
section accounts for the suppression of power at large scales
observed by WMAP, and fits the WMAP temperature two-point
correlation function for  $ 1.015 \leq\Omega_{\mathrm{tot}} \leq
1.020$ \citep{Aurich1,Aurich2}, retaining the standard
Friedmann-Lema\^{\i}tre-Robertson-Walker (FLRW) foundation for local
physics.

A preliminary search failed to find the antipodal matched circles in
the WMAP CMBR sky maps predicted for the Poincar\'e dodecahedral
space model \citep{Cornish}. In a second search, indications for
these correlated circles were found, but due to noise and foreground
structure of the CMBR maps, no final conclusion has been drawn
\citep{Aurich3}. We also note that the Doppler and integrated
Sachs-Wolfe contributions to the circles-in-the-sky are strong
enough to blur the circles, and thus the matched circles can be
overlooked in the CMBR sky maps \citep{Aurich1}. Additional effects
such as the Sunyaev-Zeldovich effect and the finite thickness of the
LSS, as well as possible systematics in the removal of the
foregrounds, can further damage the topological circle matching.

On these observational grounds, in what follows, we shall assume the
Poincar\'e dodecahedron model. 

\subsection{SNe Ia plus cosmic topology analysis}

\begin{figure}
   \centering
   \includegraphics[width=9cm]{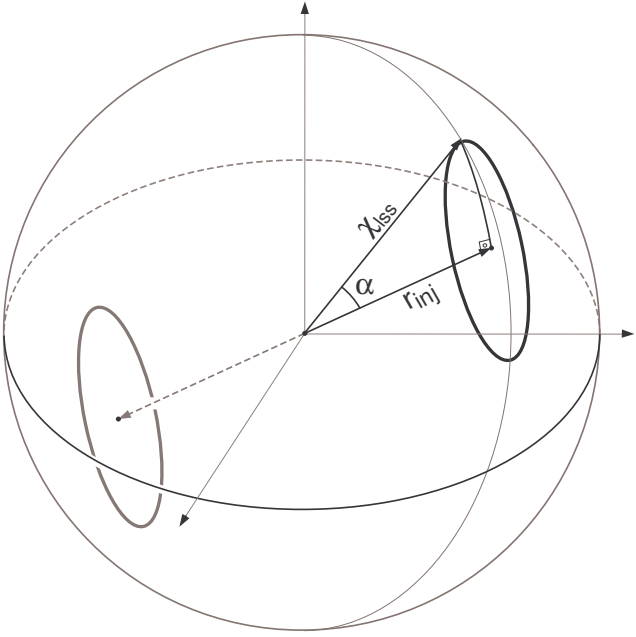}
      \caption{A schematic illustration of two antipodal
matching circles in the sphere of last scattering. The relation
between the angular radius $\alpha$ and the angular sides $r_{inj}$
and $\chi^{}_{lss}$ is given by the following Napier's rule for
spherical triangles: $\sin (\pi/2 - \alpha) = \tan r_{inj}\, \tan
(\pi/2 - \chi^{}_{lss})$ \citep[see, e.g.,][]{Coxeter}.
              }
         \label{CinTheSky}
   \end{figure}

To study the consequences of the FLRW model with the Poincar\'e
dodecahedral space section $\mathcal{D}$, we begin by recalling that
this model predicts six pairs of antipodal matched circles on the
LSS, centered in a symmetrical pattern like the faces of the
dodecahedron. Clearly the distance between the centers of each pair
of circles is twice the radius $r_{inj}$ of the sphere inscribable
in $\mathcal{D}$. Now, a straightforward use of a Napier's rule to
the right-angled spherical triangle  with elements $r_{inj}$, the
angular radius $\alpha$ of a matched circle, and the radius
$\chi^{}_{lss}$ of the last scattering sphere (see
Fig.~\ref{CinTheSky}), furnishes
\begin{equation}
\label{cosalpha} \cos \alpha = \frac{\tan r_{inj}}{\tan
\chi^{}_{lss} }\;,
\end{equation}
where  $r_{inj} = \pi/10$ for the dodecahedron. Note that
$\chi^{}_{lss}$ depends only on the cosmological scenario, and for
the $\Lambda$CDM model it reads (in units of the curvature radius)
\begin{equation}
\label{redshift-dist}  
 \chi^{}_{lss}= \sqrt{|\Omega_k|} \int_1^{1+z_{lss}} \hspace{-4mm}
\frac{dx}{\sqrt{\Omega_{m}x^3 + \Omega_kx^2 +
 \Omega_{\Lambda}}} \;,
\end{equation}
where $\Omega_k = 1-\Omega_{\mathrm{tot}}$ and $z_{lss}=1089$
\citep{WMAP-Spergel}.

Equations~(\ref{cosalpha}) and~(\ref{redshift-dist}) give the
relations between the angular radius $\alpha$ and the cosmological
density parameters $\Omega_{\Lambda}$ and $\Omega_{m}$, and thus can
be used to set bounds on these parameters. To quantify this, we
proceed in the following way. Firstly, we take the angular radius
$\alpha = 50^\circ$ estimated in \citet{Aurich1}. Secondly, we note
that measurements of the radius $\alpha$ unavoidably involve
observational uncertainties, and therefore, in order to set
constraints on the density parameters from the detection of cosmic
topology, one should take such uncertainties into account. To obtain
very conservative results, we take $\delta {\alpha} \simeq 6^\circ$,
the scale below which the circles are blurred \citep{Aurich1}.

In our statistical analysis, we use SNe Ia data from \citet{rnew}.
The total sample presented in that reference consists of 186 events
distributed over the redshift interval $0.01 \lesssim z \lesssim
1.7$ and constitutes the compilation of 
observations made by two supernova search teams plus, 16 new events
observed by the Hubble space telescope (HST). This total data set
was initially divided into ``high-confidence'' (\emph{gold}) and
``likely but not certain'' (\emph{silver}) subsets.  Here, we
consider only the 157 events that constitute the so-called
\emph{gold} sample. The confidence regions in the parametric space
$\Omega_m - \Omega_{\Lambda}$ are determined by defining a
probability distribution function ${\cal{L}} =  \int{
e^{-\chi^{2}(\mathbf{p})/2}dh}$, where $\mathbf{p}$ stands for the
parameters $\Omega_m$, $\Omega_{\Lambda}$, and $h$, and we have
marginalized over all possible values of the Hubble parameter $h$
\citep[for some recent SNe Ia analyses see][]{CP03,NP04,APSNIa04}.
The Poincar\'e dodecahedral space topology is added 
to the SNe Ia data as a Gaussian prior on the value of
$\chi^{}_{lss}$,
which can easily be obtained from 
Eqs.~(\ref{cosalpha})~--~(\ref{redshift-dist}).

\begin{figure}
   \centering
   \includegraphics[width=9cm]{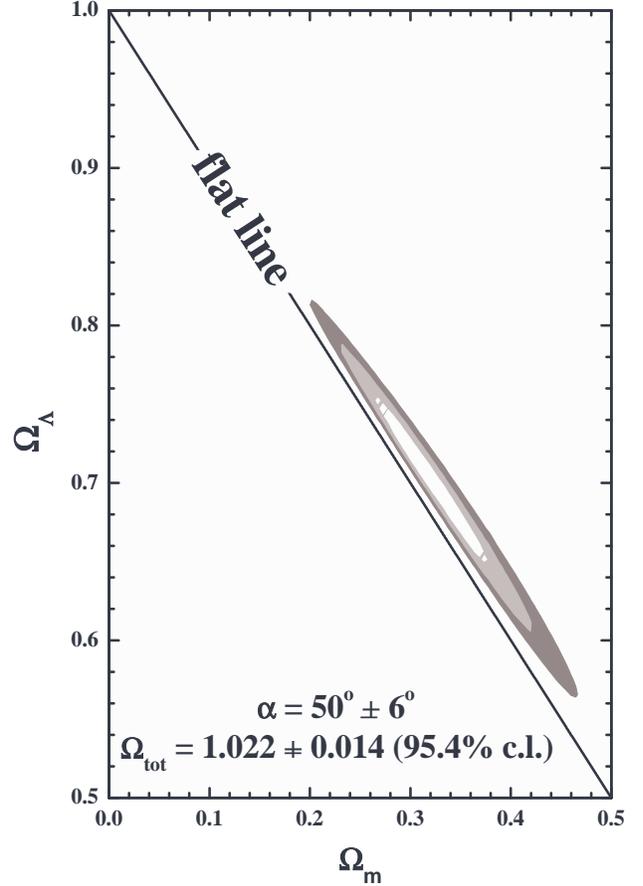}
      \caption{The 68.3\%, 95.4\%, and 99.7\% confidence
regions in the density parametric plane, which arise from the SNe Ia
plus dodecahedral space topology analysis. The best-fit values for
the dark matter and dark energy density parameters are,
respectively, $\Omega_m = 0.316^{+0.011}_{-0.009}$ and
$\Omega_{\Lambda} = 0.706^{+0.010}_{-0.009}$ at a 95.4\% confidence
level. The value of the total density parameter, as well as of the
angular radius of the circles and the corresponding uncertainties,
are also displayed.
              }
         \label{Top+SNIa}
   \end{figure}

Figure~\ref{Top+SNIa} shows the results of our joint SNe Ia plus
cosmic topology analysis. There, we display the confidence regions
(68.3\%, 95.4\%, and 99.7\%) in the parametric plane $\Omega_m -
\Omega_{\Lambda}$. Compared to the conventional SNe Ia analysis,
i.e. the one with no such cosmic topology assumption \citep[see,
e.g., Fig.~8 of][]{rnew}, it is clear that the effect of the cosmic
topology as a new cosmological observable is to considerably reduce
the area corresponding to the confidence intervals in the parametric
space $\Omega_m - \Omega_{\Lambda}$, as well as to break
degeneracies arising from the current SNe Ia measurements. The
best-fit parameters for this joint analysis are $\Omega_m = 0.316$
and $\Omega_{\Lambda} = 0.706$ with reduced $\chi^2_{min}/\nu \simeq
1.13$ ($\nu$ is defined as degrees of freedom). At a 95.4\%
confidence level (c.l.) we found $\Omega_m =
0.316^{+0.010}_{-0.009}$ and $\Omega_{\Lambda} = 0.706 \pm 0.010$,
which corresponds to $\Omega_{\mathrm{tot}} = 1.022 \pm 0.014$. Note
that this value of the total energy density parameter derived from
our SNe Ia plus topology statistics is in full agreement with those
reported by the WMAP team, $\Omega_{\mathrm{tot}} = 1.02 \pm 0.02$
\citep{WMAP-Spergel}, as well as with the value obtained by fitting
the Poincar\'e dodecahedral power spectrum for low multipoles 
with the WMAP data, i.e. $ 1.015 \leq \Omega_{\mathrm{tot}} \leq
1.020$ \citep{Aurich1} and $\Omega_{\mathrm{tot}}\simeq 1.013$
\citep{Poincare}.

Concerning the above analysis it is also worth emphasizing three
important aspects at this point. First, the range $1.015 \leq
\Omega_{\mathrm{tot}} \leq 1.020$ in which the Poincar\'e
dodecahedral space model fits the WMAP data (and also gives rise to
six pairs of matching circles) has not been used as a prior of our
statistical data analysis. Second, the best-fit values for both
$\Omega_m$ and $\Omega_{\Lambda}$ (and, consequently, for
$\Omega_{\mathrm{tot}}$) depend very weakly on the value used for
the angular radius $\alpha$ of the circle. As an example, by
assuming $\alpha = 11^\circ \pm 1^\circ$, as suggested in
\citet{Roukema}, it is found that $\Omega_m = 0.312^{+
0.078}_{-0.072}$, $\Omega_{\Lambda} = 0.698^{ + 0.072}_{ - 0.078}$,
and $\Omega_{\mathrm{tot}}= 1.010\pm0.002$
at a 95.4\% (c.l.), which is very close to the value found by
considering $\alpha = 50^\circ$ \citep{Aurich1} with an uncertainty
of $6^\circ$. Third, the uncertainty on the value of the radius
$\alpha$ alters the width corresponding to the confidence regions,
without having a significant effect on the best-fit values. Finally,
we also notice that, by imposing the topological prior, the
estimated value for the matter density parameter is surprisingly
close to those suggested by dynamic or clustering estimates
\citep[see, e.g.,][]{calb,DBW1997,wm1,wm,Pope2004}. On the other
hand, as shown in \citet{rnew} \citep[see
also][]{CP03,NP04,APSNIa04}, the conventional SNe Ia analysis
(without the above cosmic topology constraint) provides $\Omega_m
\simeq 0.46$, which is $\sim 1\sigma$ off from the central value
obtained by using independent methods, as for instance, the mean
relative peculiar velocity measurements for pairs of
galaxies~\citep{wm1}.

\section{Final remarks}

Fundamental questions, such as whether the Universe will expand
forever or eventually re-collapse and what are its shape and size,
are associated with the nature of its constituents as well as with
the measurements of both the local curvature and the global topology
of the $3$-dimensional world. The so-called ``circles-in-the-sky"
method makes it apparent that a non-trivial detectable topology of
the spatial section can be probed for any locally homogeneous and
isotropic Universe, with no assumption about the cosmological
density parameters. In this article, we have shown that the
knowledge of spatial topology of the Universe not only dictates the
sign of
its local curvature 
(and therefore its geometry), but also imposes very restrictive
constraints on the density parameters associated with dark matter
($\Omega_m$) and dark energy ($\Omega_{\Lambda}$). Indeed, by
combining the detection of the cosmic topology through the
``circles-in-the-sky" method with the current SNe Ia observations,
we have shown that the effect of the cosmic topology as a
cosmological observable is to drastically reduce the degeneracies
inherent to current SNe data, providing limits on the cosmological
density parameters, which cannot presently be obtained from
combinations of the current cosmological data. This role of cosmic
topology has previously been emphasized in the context of cosmic
crystallography by \citet{Uzan}. We underline the fact that the-best
fit values are not the most important outcome of our work, since the
dodecahedral space model has not been confirmed as the ultimate
global topology of the Universe.

We emphasize that even though the precise value of the radius
$\alpha$ of the circle and its uncertainty (fundamental quantities
in our analysis) can be modified by more accurate analysis and
future observations, the general aspects of our analysis remain
essentially unchanged, since the best-fit values of the cosmological
parameters depend very weakly on $\alpha$, and the value of
uncertainty $\delta \alpha$ primarily alters the confidence
uncertainty area in the density parametric plane $\Omega_m -
\Omega_{\Lambda}$. On the other hand, regarding the possibility of
using the observational results to guide the search for the circles
in the sky, from a SDSS plus WMAP combination of large-scale
structure, SNe Ia, and CMBR data \citep{Tegmark_et_al}, we can only
place an upper bound on the angular radii of the circles for a
Poincar\'e dodecahedral topology, namely $\alpha < 70^\circ$, which
is consistent with value of $\alpha$ we have used in this work.

Given the immense efforts expended in the quest for the local
curvature of the Universe, we believe that our results reinforce the
cosmological interest in the search for definitive observational
evidences of a non-trivial cosmic topology. Further investigations
of the other globally homogeneous spherical spaces that also fit
current CMBR data are in progress and will be presented in a
forthcoming article.

\emph{Acknowledgements.} The authors are grateful to A.F.F. Teixeira
for valuable discussions. We thank CNPq for the grants under which
this work was carried out. JSA is also supported by Funda\c{c}\~ao
de Amparo \`a Pesquisa do Estado do Rio de Janeiro (FAPERJ).

\end{document}